\documentclass{article}
\usepackage{amsmath,amssymb}
\usepackage{cite}
\usepackage{color}
\usepackage[utf8]{inputenc}
\usepackage{graphicx}
\usepackage{authblk}
\title{Generation of megatesla magnetic fields by intense-laser$-$driven microtube implosions}
\author[1]{M. Murakami$^\ast$}
\author[2]{J.J. Honrubia}
\author[3]{K. Weichman}
\author[3]{A.V. Arefiev}
\author[4,5]{S.V. Bulanov}
\affil[1]{Institute of Laser Engineering, Osaka University, Suita, Osaka 565-0871, Japan}
\affil[2]{ETSI Aeron\'{a}utica y del Espacio, Universidad Polit\'{e}cnica de Madrid, Madrid, Spain}
\affil[3]{University of California San Diego, 9500 Gilman Drive, La Jolla, California 92093-0411, USA}
\affil[4]{Institute of Physics of the ASCR, ELI-Beamlines, Na Slovance 2, 18221, Prague, Czech Republic}
\affil[5]{Kansai Photon Science Institute, National Institutes for Quantum and Radiological Science and Technology, Kizugawa, Kyoto 619-0215, Japan}

\begin{document}

\maketitle

\begin{abstract}
A microtube implosion driven by ultraintense laser pulses is used to produce ultrahigh magnetic fields.
Due to the laser-produced hot electrons with energies of  mega-electron volts, cold ions in the inner wall surface implode towards the central axis.
By pre-seeding uniform magnetic fields on the  kilotesla order, the Lorenz force induces the Larmor gyromotion of the imploding ions and electrons. Due to the resultant collective motion of relativistic charged particles around the central axis, strong spin current densities of $\sim$ peta-ampere/cm$^{2}$ are produced with a few tens of nm size,  generating megatesla-order magnetic fields.
The underlying physics and important scaling are revealed by particle simulations and a simple analytical model. The concept holds promise to open new frontiers in many branches of fundamental physics and applications in terms of ultrahigh magnetic fields.

\end{abstract}

\section*{Introduction}
Laboratory generation of strong magnetic fields have been intensively studied \cite{Sakh65,Fowl60,Cnar66, Daid86,
Felb88, Veli85, Miur96, Pego97, Cour05, Gotc09, Knau10, Fuji13, Sant15, Tikh17, Naka18, Lecz18, Wang19, Park19}, because such fields may realize new experimental tools for fundamental studies and  support diverse applications. 
Examples include 
plasma and beam physics \cite{Sagd66, Aref16, Star16, Bail18, Nakat18, Weng17}, 
astro- \cite{Sant99, Gour14} and solar-physics \cite{Kopp76, Masu94}, 
atomic and molecular physics \cite{Gilc98}, and materials science \cite{Lai01, Herl03}.
Magnetic field reconnection \cite{Kopp76, Masu94}, generation of collisionless shock \cite{Sagd66}, gamma-ray and pair production \cite{Ribe16, Jans18, Koga19},  and fusion application in a strongly magnetized plasma \cite{Slut12, Wang15, Honr17, Sant18} are receiving increased attention.
{\color{black}Although laser$-$solid interactions numerically predict
 magnetic fields $\lesssim$ a few hundreds kT \cite{Star16, Nakat18, Park19},
the highest magnetic field  experimentally observed  to date is on the kilotesla (kT)  order \cite{Fuji13,  Naka18}.} 

\begin{figure}
\centering
\includegraphics[width=120mm]{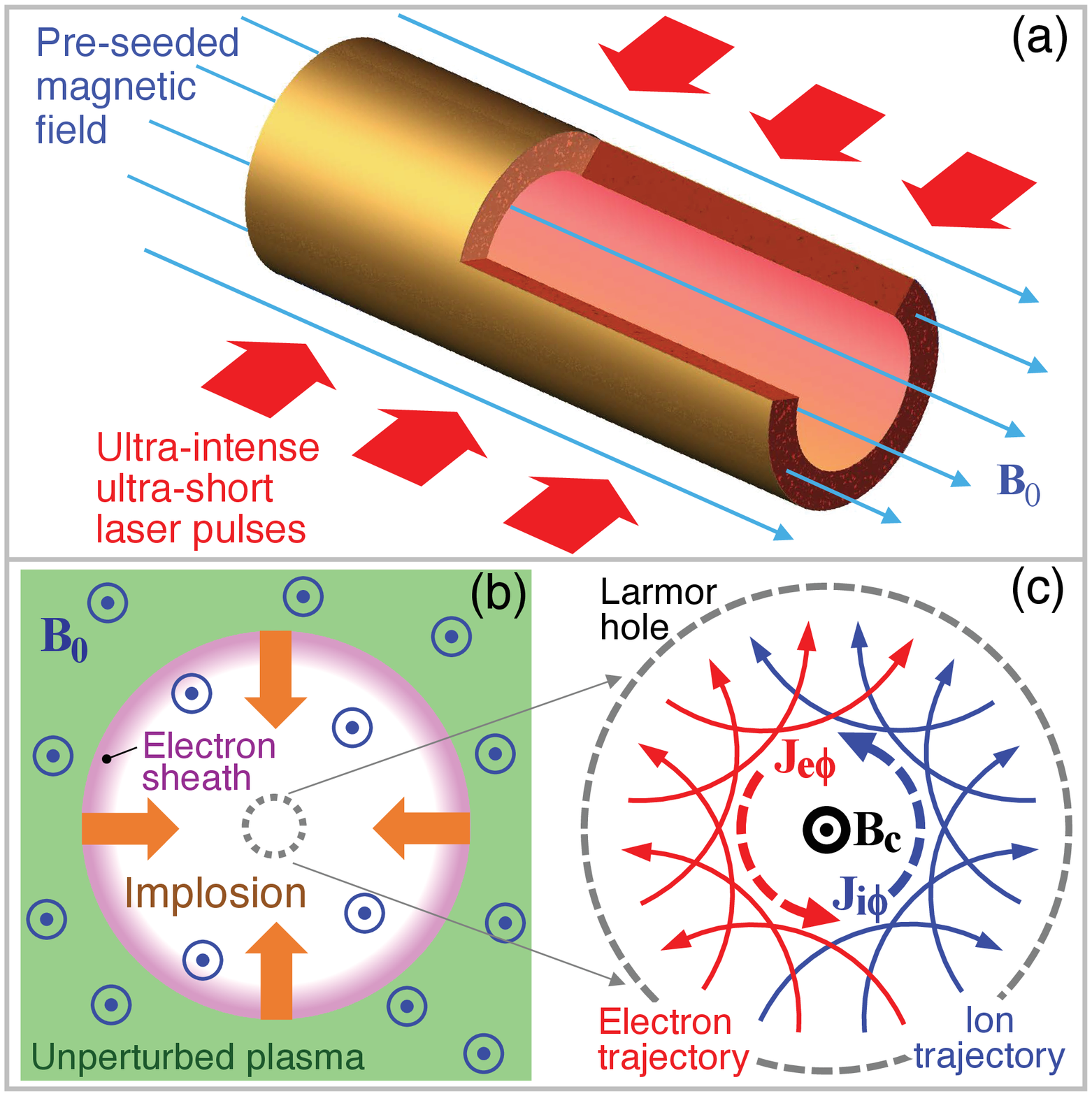}
\caption{(a) Perspective view of a microtube irradiated by ultraintense laser pulses (laser configuration is just schematic). Uniform external magnetic field ${\bf B}_0$ is pre-seeded prior to  main laser illumination. (b) Top view of the inner plasma dynamics. Laser-produced hot electrons drive isothermal expansion of the inner-wall plasma into vacuum. (c) Ultrahigh magnetic field ${\bf B}_{\rm c}$ is generated at the center due to the collectively formed currents by ions and electrons, which are deflected in opposite directions by ${\bf B}_0$.}
\label{fig1}
\end{figure}

Here we propose a novel concept called a microtube implosion (MTI), which generates megatesla (MT) magnetic fields utilizing a structured target and intense laser pulses.
Suppose that a long-stretched cylindrical target contains a coaxial hollow cylindrical space with an inner radius of  $R_0\sim 1-10$ $\mu$m (Fig. \ref{fig1}a). 
Irradiating the target by ultraintense femtosecond laser pulses with an intensity of $I_{\rm L} \sim 10^{19}-10^{22}$ Wcm$^{-2}$ generates
 hot electrons with temperatures of $T_{\rm e}\sim 1-$ a few 10's mega-electron volts (MeV) according to the ponderomotive scaling \cite{Wilk01}.
 {\color{black}Note that the ponderomotive scaling  does not resemble the true dynamics of electrons on a solid surface with a steep density gradient, where the prerequisites for decoupling the quiver and envelope motion of electrons do not hold.}
Hot electrons ionize the target material to produce a plasma of atomic mass number $A$ and  initial ion density $n_{\rm i0}\sim 10^{23}$ cm$^{-3}$  to  ionization state $Z$. 

The hot electrons are so energetic that some of them exit the target wall and enter the cavity. Therein the electron pressure and the electrostatic force balance with each other to form  an electron sheath on the plasma/vacuum interfaces. The surface ions are accelerated inward (implosion, Fig. 1b) through expansion into a vacuum by the sheath electric field  \cite{Norr99, Wilk92, Mora03, Fuch06,Mura06}. 
In the ideal situation where a system has a perfect axial symmetry, the temporal evolution of the imploding and exploding plasma should also be axially symmetric. In this configuration, a magnetic field does not evolve. However, if
a pre-seeded magnetic field is introduced into the system,  an extraordinary magnetic field can be generated at the center with a  $2-3$  orders of magnitude larger magnification factor.

A uniform magnetic field ${\textbf{\textit B}}_0$ on the kT order, which is parallel to the cylindrical axis ($z$-axis),
{\color{black} is pre-seeded by an external laser-plasma device such as a capacitor coil  \cite{Daid86, Fuji13, Sant15, Tikh17}. 
 Using a ns-long laser,  such a seed field quickly rises on the sub-ns time scale and diffuses into the MTI target nearly simultaneously, 
 and then slowly decays on time scales $\gtrsim$ 10 ns, which are characterized by impedance of the capacitor-coil.}
Thus, the lifetime  of such a pre-seeded magnetic field $\gtrsim 10$ ns is much longer than the characteristic time scale of MTI $\sim$ 100 fs. During the implosion, the Lorenz force deflects ions and electrons clockwise and anticlockwise, respectively, gaining azimuthal momentum, as depicted in Fig. \ref{fig1}c. 
The ion trajectories draw circles with Larmor radii $\sim 0.1 - 1$ mm for typical laser and target parameters in MTI. 
In particular, the envelope of the ion paths forms a nanometer-scale hole at the center (hereafter called the ``Larmor hole").
Since electrons are negatively charged, the resultant direction of the electron current ${\textbf{\textit J}}_{\rm e\phi}$ is anticlockwise, which is the same as that of the ion current ${\textbf{\textit J}}_{\rm i\phi}$. Then ultraintense spin currents on the order of $10^{15}$ Acm$^{-2}$ run around the Larmor hole. Consequently, the currents from the ions and  electrons work together to generate MT-order magnetic field ${\textbf{\textit B}}_{\rm c}$ at the center.

{\color{black}Compared with other conventional approaches, the most innovative point of the current concept lies in the geometrically unique plasma flow.
A cylindrically converging flow composed of relativistic electrons and ions, which are infinitesimally twisted by the pre-seeded magnetic field in opposite  directions, can effectively produce  ultrahigh spin currents and consequently, ultrahigh magnetic fields. In addition, the current geometry may be  better suited for many practical purposes.
}
 
For over 50 years, researchers have strived to realize  high magnetic fields.
Many approaches have been employed, including high explosives \cite{Sakh65,Fowl60}, electromagnetic implosions \cite{Cnar66, Miur96}, high-power lasers \cite{Gotc09, Knau10}, and Z pinches \cite{Felb88, Veli85}.
The principal physical mechanism of these works is based on  magnetic flux compression (MFC) using hollow cylindrical structures and pre-seeded magnetic fields.
The present scheme also uses a similar physical configuration.
However, MTI differs from MFC because 
the ultrahigh magnetic fields in MTI are generated by the spin currents induced by collective Larmor gyromotions. 

 
\section*{Results}

\subsection*{Two-dimensional particle simulation}
To demonstrate the expected behavior of MTI, we perform 2D $(x, y)$ PIC simulations using the open-source fully relativistic code EPOCH \cite{Arbe15}.
{\color{black}In this first part of EPOCH simulations, we employ rather simple and ideal physical conditions to effectively extract the salient features of the underlying MTI physics.}
First, the simulation uses the periodic boundary conditions for particles and fields,  where the hollow cylindrical volume is placed at the middle of the square computational domain. 
This configuration simulates  collective targets with multiple equally spaced microtubes inside a heated material. We set the lengths for the unit cell size and full span of each side of the square domain as 6.25 nm and 10 $\mu$m, respectively. Therefore, the whole computational domain size is $1600\times 1600$ mesh$^2$.  The initial inner radius of the microtube is $R_0 =$ 3  $\mu$m. 

Second, since hot-electron average energy $\mathcal{E}_{\rm he.av}$ spans the relativistic regime for the parameters of interest, we use the Maxwell-J{\" u}ttner (M-J) distribution \cite{Jutt11} rather than the Maxwell-Boltzmann (M-B) distribution for the non-relativistic regime.
{\color{black}The M-J distribution defines the hot electron population in terms of the Lorenz factor $\gamma$ as $f(\gamma)=\frac{\gamma^2 \beta}{\Theta K_2(1/\Theta)}\exp(-\frac{\gamma}{\Theta})$, where $\beta=v/c =\sqrt{1-1/\gamma^2}$ and $\Theta=T_{\rm e}/m_{\rm e}c^2$ with 
$T_{\rm e}$, $m_{\rm e}$, and $c$ being  the electron temperature, the electron rest mass, and the speed of light, respectively;
 $K_2$ is the modified Bessel function of the second kind.}
The relation between $\mathcal{E}_{\rm he.av}$ and $T_{\rm e}$  significantly differs between
the two distributions. That is, $\mathcal{E}_{\rm he.av}=\frac32 T_{\rm e}$  for the M-B distribution ($\mathcal{E}_{\rm he.av}\ll m_{\rm e}c^2$) while 
$\mathcal{E}_{\rm he.av}\simeq 3T_{\rm e}-m_{\rm e}c^2$ for the M-J distribution ($\mathcal{E}_{\rm he.av}\gg m_{\rm e}c^2$).  

{\color{black}
It should be noted that on such an ultrashort timescale as femtoseconds, there is insufficient time for electrons to be thermalized \cite{Bezz80, Klug18}. In this sense, employing the M-J distribution, which is characterized by a specific temperature, may not be legitimate. However,  high-energy-tail electrons, whose population decreases  exponentially with energy, predominantly influence the energy transport and thus the dynamics of the overall system  \cite{Mura06}. In fact, both the M-J and M-B distributions have such an exponential dependence in their functional forms. For this reason, employing the M-J distribution is an acceptable choice.
Actually, simulations have confirmed  that the same value of $\mathcal{E}_{\rm he.av}$ yields a similar result for the implosion dynamics and the generation of the magnetic field for  both  energy distributions. Therefore, $\mathcal{E}_{\rm he.av}$ rather than $T_{\rm e}$ is employed below as a principal parameter.
Note that we later provide another set of simulation results as a proof-of-principle,
using more practical conditions that take the laser$-$matter interactions into account, where the electron population is not approximated by the M-J distribution.}

Figure \ref{fig2} shows the temporal evolution of the normalized densities of ions, $\tilde{n}_{\rm i}=n_{\rm i}/n_{\rm i0}$, and electrons, $\tilde{n}_{\rm e}=n_{\rm e}/n_{\rm e0}$, under $n_{\rm e0}=Z n_{\rm i0}$  for a fully ionized carbon plasma with $A=12$ and $Z=6$.
The solid and dashed curves indicate the EPOCH results and the model prediction, respectively. 
The model is described later.
Initially,  the inside of the  tube is empty, and the remaining volume is filled with uniform  ions with  $T_{\rm i}=10$ eV and $n_{\rm i0}=1\times 10^{23}$ cm$^{-3}$  and uniform electrons with  $\mathcal{E}_{\rm he.av}=5$ MeV.
The pre-seeded magnetic field $B_0=4$ kT  is distributed uniformly over the entire computational domain.

After launching the plasma expansion into a vacuum at $\tau=0$, the implosion phase is observed for a period, $\tau\lesssim 70$ fs (Fig. \ref{fig2}). The implosion velocity of the innermost ions remains nearly constant at $v_{\rm i}\simeq 6\times 10^9$ cm/s before the cavity collapse. Macroscopically, ions and electrons in the imploding plasma layer move together and maintain  charge neutrality.
The electron sheath thickness at the plasma/vacuum interface is roughly equal to the local electron Debye length $\lambda_{\rm De}=(T_{\rm e}/4\pi n_{\rm e} e^2)^{1/2}\sim 150$ nm, where $e$ denotes the elementary charge,  $T_{\rm e}\approx 1.8$ MeV  ($\mathcal{E}_{\rm he.av} = 5$ MeV with the M-J distribution), and $n_{\rm e}\approx 6\times 10^{21}$ cm$^{-3}$ (Fig. \ref{fig2}). 

Upon cavity collapse, the head group of imploding ions passes the target center at the Larmor hole radius $r=R_{\rm H}$ and expands outward. 
The mean-free-path of ion-ion collisions is roughly given by $\ell_{\rm ii}\sim T_{\rm e}^2/4\pi n_{\rm i} Z^2 e^4$, which amounts to $\sim$ 4 cm $\gg R_0$ under $T_{\rm e}=2$ MeV, $Z=6$, and $n_{\rm i}=10^{23}$ cm$^{-3}$.
Hence, these ions collisionlessly intersect other ions, which are still  imploding toward the center. Meanwhile, the central density increases to the same order as its initial value
due to the geometrical accumulation effect (Fig. \ref{fig2}, inset). 
The Larmor hole radius, $R_{\rm H}\approx 20$ nm, indicated at the top of the inset corresponds to the analytical prediction at $\tau=90$ fs [Eq.(\ref{RH=})]. The Larmor hole is also seen in the simulation result as the one-humped structure, but the simulated one grows more quickly than the model and expands outward.
This is attributed to the fact that a highly compressed ion sphere is created at the center and the strong electrostatic field radially pushes the ions outward.  

Figure \ref{fig3} shows snapshots taken from the dominant period of the magnetic field generation, $\tau\approx 80 - 120$ fs: (a) the normalized ion density $\tilde{n}_{\rm i}=n_{\rm i}/n_{\rm i0}$, (b) the normalized electron density $\tilde{n}_{\rm e}=n_{\rm e}/n_{\rm e0}$, (c) the azimuthal electron current $J_{\rm e\phi}$, and (d) the magnetic field $B_{\rm z}$. 
Comparing Figs. 3a and 3b provides insight on how the core plasma develops. The one-humped structure forms in the central region and  oscillates at the ion-plasma frequency $\omega_{\rm pi}=(4\pi n_{\rm i}Z^2e^2/m_{\rm i})^{1/2}$ to emit  compression waves outward at sound speed $c_{\rm s}=(Z T_{\rm e}/m_{\rm i})^{1/2}$. Applying the numbers used in Fig. \ref{fig3} (i.e., $n_{\rm i}=1\times 10^{23}$ cm$^{-3}$ and $T_{\rm e}=1.8$ MeV) yields $c_{\rm s}\simeq 9\times 10^8$ cm/s and the cycle $\tau_{\rm cyc}=2\pi/\omega_{\rm pi}\simeq 9$ fs ($\nu\equiv \tau^{-1}_{\rm cyc}\simeq110$ THz), which agree well with the simulation result.

According to Ampere's law, $c\nabla\times{\textbf{\textit B}}=4\pi{\textit{\textbf J}}+\dot{\textbf{\textit E}}$, the azimuthal current distribution, $J_\phi=J_{\rm e\phi}+J_{\rm i\phi}$, directly contributes to the magnetic field $B_{\rm c}$ generated at the center.
The azimuthal electron current density $J_{\rm e\phi}$ dynamically evolves around the center over the distance approximately equal to the local Debye length $\lambda_{\rm De}\sim 100-150$ nm (Fig. \ref{fig3}c).
According to Faraday's law, $c\nabla\times {\textit{\textbf E}}=-\dot{\textbf{\textit B}}$,
when $B_{\rm c}$ reaches its peak, the displacement current ($\propto \partial E_\phi/\partial t$) becomes substantially small. Then, $B_{\rm c}$ is given as the sum, $B_{\rm c}=B_{\rm ce}+B_{\rm ci}$, where 
 $B_{\rm ce}=(4\pi/c)\int^\infty_0 J_{\rm e\phi}{\rm d}r$  and $B_{\rm ci}=(4\pi/c)\int^\infty_0 J_{\rm i\phi} {\rm d}r$ are the contributions from electrons and ions, respectively.
Due to the high mobility of electrons, the effect of electron currents on the magnetic field generation dominates over that of ion currents.
MTI simulations indicate that $B_{\rm ce}/B_{\rm ci}\sim 3-4$ or equivalently 
$B_{\rm c}/B_{\rm ci}\sim 4-5$ is kept nearly constant. For example,  $B_{\rm c}\simeq 0.95$ MT and $B_{\rm ce}\simeq 0.71$ MT are derived at $\tau=105$ fs from Fig. \ref{fig3}c,d, which correspond to $B_{\rm c}\simeq 4 B_{\rm ci}$.

\subsection*{Model}
Here,  we describe the ion dynamics 
in terms of a semi-analytical model and
demonstrate that it forms the basis of the whole system.
The time origin matters when comparing the model to the simulation results. To avoid confusion, hereafter, we employ the time variable, $t$, instead of $\tau$ used for the simulation. 
Suppose that a planar plasma is held at rest in a half-infinitely stretched region $-\infty<x\le 0$ for $t\le 0$, which is composed of uniform cold ions and hot electrons with densities $n_{\rm i}$ and $n_{\rm e}$, respectively. We  postulate that the plasma is charge neutral, i.e., $Zn_{\rm i}=n_{\rm e}$. In addition, hot electron temperature $T_{\rm e}$ is assumed to be constant  both spatially and temporally due to the high conductivity. 
Once the boundary between the vacuum and plasma is set free at $t=0$, the plasma begins to expand into the vacuum.
The ion motion is governed by the following hydrodynamic system describing the mass and momentum conservation as
\begin{eqnarray}
\label{BE1}&&
\frac{\partial n_{\rm i}}{\partial t}+ \frac{\partial}{\partial x} (n_{\rm i} v_{\rm i}) = 0,\\ 
\label{BE2}&&
\frac{\partial v_{\rm i}}{\partial t} + v_{\rm i} \frac{\partial v_{\rm i}}{\partial x} = -\frac{c_{\rm s}^2}{n_{\rm i}}\frac{\partial n_{\rm i}}{\partial x},
\end{eqnarray}
where $n_{\rm i}(x, t)$ and $v_{\rm i}(x, t)$ are the number density and the velocity of the ions, respectively.
Grevich {\it et al.} \cite{Gure65} found a self-similar solution to the above system, where the physical quantities are expressed in terms of  a single dimensionless coordinate defined by $\xi = x/c_{\rm s}t \,(\ge -1)$ in the forms of
$n_{\rm i}=n_{\rm i0} {\rm e}^{-\xi -1}$ 
and $v_{\rm i}= (\xi +1)c_{\rm s}$.
Under the self-similar solution, the plasma expands to the right supersonically for $x>0$ or $\xi>0$, while the rarefaction wave propagates to the left at the sound speed $c_{\rm s}$, corresponding to the path, $\xi=-1$.

The two physical ingredients, collisionless ions and isothermal electrons, provide insight to harness Grevich's self-similar solution {\color{black} as a useful approximation} and to describe the kinetic behavior of the ions.
To accomplish this, a geometrical modification needs to be added to the self-similar solution.
The resultant system behaves such that individual fluid elements can penetrate each other in  cylindrically converging and diverging processes.

Suppose that an ion with mass $m_{\rm i}$ and ionization state $Z$ is moving at a constant speed $v_{\rm i}$ on the $xy$-plane in a uniform magnetic field ${\textbf {\textit B}}_0$, which is parallel to the $z$-axis. The ion draws a circular orbit with a Larmor radius $R_{\rm L}=m_{\rm i}v_{\rm i}c/ZeB_0$, where $B_0=|{\textbf {\textit B}}_0|$. If the position and velocity of the ion are specified at $t=0$ to be $(x,y)=(R_0,0)$ and $(\dot{x},\dot{y})=(-v_{\rm i},0)$, respectively, the ion moves on the circle, $(x-R_0)^2+(y-R_{\rm L})^2=R_{\rm L}^2$.

Here, it is useful to introduce cylindrical coordinates, $r=\sqrt{x^2+y^2}$ and $\phi=\tan^{-1}(y/x)$.
The ion path $s$ along its Larmor circle is measured with the distance from the initial point $(x,y)=(R_0,0)$ or with the polar angle $\theta=\sin^{-1}[(R_0-x)/R_{\rm L}]$ pivoting around the guiding center $(x,y)=(R_0, R_{\rm L})$, as illustrated in Fig. \ref{fig4}a.
It should be noted that Fig. \ref{fig4}a is not to scale.
The dimensionless coordinate of the self-similar solution is then redefined in terms of $s$ and $\theta$ as 
$\xi=s/c_{\rm s}t=  R_{\rm L}\theta/c_{\rm s}t$.
Note that $v_{\rm i}$ and $R_{\rm L}$ are functions of $\xi$.
Consequently, Grevich's self-similar solution for a planar system is reformed for a cylindrical system as
\begin{eqnarray}
\label{SSM1}&&
n_{\rm i}=n_{\rm i0}\frac{R_0}{r}{\rm e}^{-\xi -1}, \\
\label{SSM2}&&
 \left(v_{\rm ir},v_{\rm i\phi}\right) 
=\frac{(\xi+1)c_{\rm s}}{r}
\left(\pm \sqrt{r^2-R_{\rm H}^2},\, R_{\rm H}\right),
\end{eqnarray}
where $v_{\rm ir}$ and $v_{\rm i\phi}$ denote the radial and azimuthal component of the ion velocity, respectively.
{\color{black}The reformed set, Eqs. (3) and (4), rigidly satisfies the mass conservation law for a cylindrical geometry such that 
the factor $R_0/r$ in Eq. (\ref{SSM1})  explains the geometrical accumulation effect.}
The  minus and plus signs of the double sign in Eq. (\ref{SSM2}) correspond to the converging and diverging phases, respectively.
The relation between $v_{\rm ir}$ and $v_{\rm i\phi}$ can be understood by considering the simplified physical picture of a single fast ion approaching the center along a straight line, $y=R_{\rm H}$, with a constant speed $v_0$ at $r=\infty$, i.e., $(v_{\rm ir}, v_{\rm i\phi})=(-v_0, 0)$. The ion passes  the origin $(x, y)=(0, 0)$ at the shortest distance  $r=R_{\rm H}$, when the velocities replace each other, i.e., $(v_{\rm ir}, v_{\rm i\phi})=(0, v_0)$.
Therefore, the higher the implosion velocity of an ion, the higher the current and resultant magnetic fields around the center, i.e., $B_{\rm z}\propto J_{\rm i\phi}=Zev_0 R_{\rm H}/r$. 

\begin{figure}
\centering
\includegraphics[width=120mm]{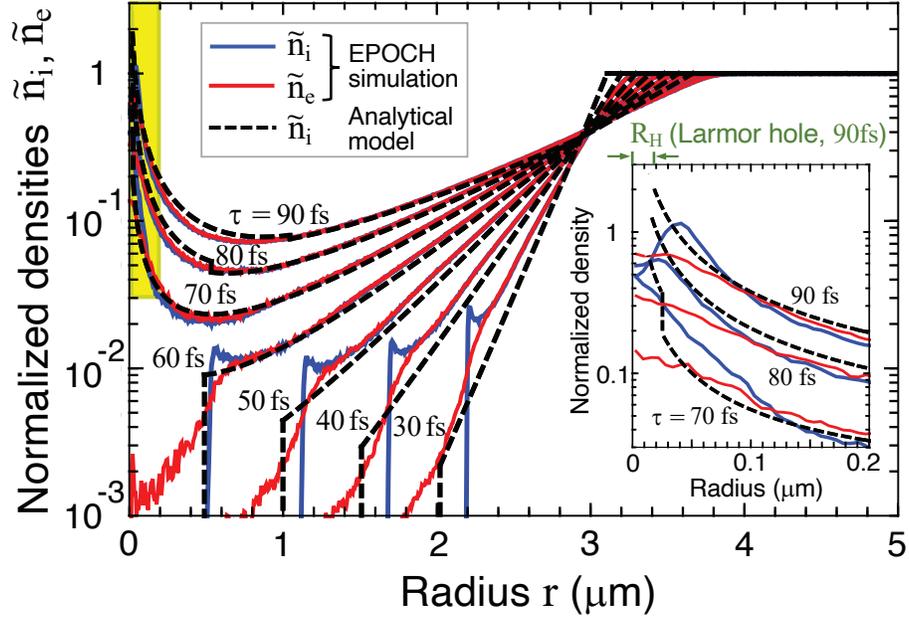}
\caption{Temporal evolution of the normalized densities of ions $\tilde{n}_{\rm i}=n_{\rm i}/n_{\rm i0}$ and electrons $\tilde{n}_{\rm e}=n_{\rm e}/n_{\rm e0}$ under charge neutrality $n_{\rm e0}=Z n_{\rm i0}$ with $Z=6$ (fully ionized carbon plasma). Other fixed parameters are $R_0=3\,\mu$m, $n_{\rm i0}=1\times 10^{23}$ cm$^{-3}$ (i.e., $2$ g$\,$cm$^{-3}$), $B_0=4$ kT, and $\mathcal{E}_{\rm he.av}=5$ MeV. Dashed curves are obtained by the model lines for constant velocities (Fig. \ref{fig4}b) and the density given by Eq.(\ref{SSM1}).
Inset shows a magnified view of the yellow-painted area. }
\label{fig2}
\end{figure}

\begin{figure}
\centering
\includegraphics[width=120mm]{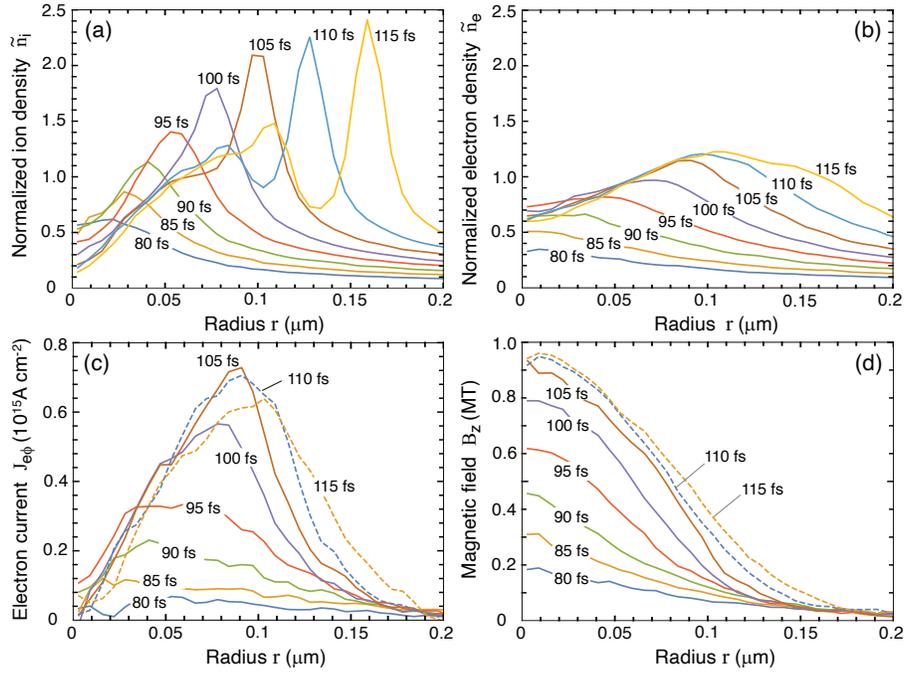}
\caption{Snapshots taken from the dominant period for the magnetic field generation, $\tau\approx 80 - 120$ fs. Fixed parameters are the same as those in Fig. \ref{fig2}. (a) Normalized ion density $\tilde{n}_{\rm i}=n_{\rm i}/n_{\rm i0}$, (b) normalized electron density $\tilde{n}_{\rm e}=n_{\rm e}/n_{\rm e0}$, (c) azimuthal electron current $J_{\rm e\phi}$, and (d) magnetic field $B_{\rm z}$. 
}
\label{fig3}
\end{figure}

\begin{figure}
\centering
\includegraphics[width=120mm]{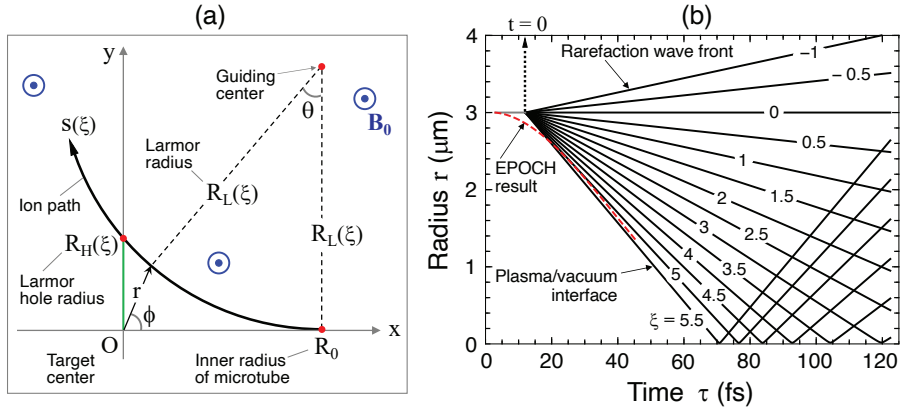}
\caption{Analytical model: (a)Schematic explaining the relation between the key parameters (not to scale).  Curve with its coordinate $s$ stands for an ion path, on which the ion implodes at a constant speed as a function of  $\xi$. In most practical cases,  $R_{\rm H} ({\rm nm}) \ll R_0 (\mu {\rm m}) \ll R_{\rm L} ({\rm mm})$.
(b)Radius$-$time diagram of $\xi$-contour lines under the same parameters as in Figs. 2 and 3. Temporal evolution of the density profile shown in Fig. \ref{fig2} is directly obtained from this diagram coupled with Eq. (\ref{SSM1}). The time lag of 12 fs between $\tau=0$ and $t=0$ is fixed such that the timing of the cavity collapse coincide for both the simulation and the model.
}
\label{fig4}
\end{figure}

Although Grevich's solution gives a simple  physical picture of  plasma expansion into a vacuum, it lacks important information such as the location of ion front $x_{\rm f}$.
In fact, Grevich's solution gives an infinite propagation speed of the ion front,  i.e., $\dot{x}_{\rm f}\rightarrow\infty$ as $\xi\rightarrow\infty$.
Assuming that the plasma expands adiabatically, a reasonable approximation for the dimensionless coordinate of the ion front $\xi_{\rm f}$ is obtained such that the plasma front expands at the speed $\dot{x}_{\rm f}=2(\gamma-1)^{-1}c_{\rm s}$ \cite{Land59}, where $\gamma$ denotes the adiabatic index. Meanwhile, the self-similar solution based on the isothermal assumption gives the speed of a fluid element at $\xi=\xi_{\rm f}$ as  $\dot{x}_{\rm f}=(\xi_{\rm f}+1)c_{\rm s}$. Equating the two speeds gives $\xi_{\rm f}=(3-\gamma)/(\gamma-1)$. In particular, the adiabatic index for the relativistic electrons is $\gamma=4/3$ \cite{Land59}, yielding $\xi_{\rm f}=5$. This result well explains $\xi_{\rm f}\simeq 5.5$ obtained from the simulation (Figs. \ref{fig2} and \ref{fig4}b).

Figure \ref{fig4}b shows the $r-t$ diagram  obtained 
from the reformed self-similar system, 
under the conditions in  Figs. \ref{fig2} and \ref{fig3}, where the curves correspond to different values of $\xi$. The time origin of the model corresponding to $\tau=12$ fs is chosen such that the cavity-collapse timings coincide with each other on the horizontal axes. This can be confirmed by
the red dashed curve, which shows the trajectory of the innermost ions at an early stage of implosion according to  the EPOCH simulation (Fig. \ref{fig2}). Combining the curves in Fig. \ref{fig4}b and the density profile given by Eq. (\ref{SSM1}) leads to the dashed curves in Fig. \ref{fig2} as the model predictions. 

The Larmor hole radius $R_{\rm H}$ is obtained from the geometrical consideration under $R_{\rm H}\ll R_0\ll R_{\rm L}$ to be $R_{\rm H}\simeq R_0^2/2R_{\rm L}$,  which is explicitly rewritten as
\begin{equation}\label{RH=}
\frac{R_{\rm H}(t)}{1{\rm nm}}\simeq 
43 \frac{Z}{A} \left(\frac{B_0}{1{\rm kT}}\right) \left(\frac{R_0}{3\mu{\rm m}}\right)^2 \left( \frac{(\xi_{\rm c}(t)+1)c_{\rm s}}{10^9 {\rm cm/s}}\right)^{-1},
\end{equation}
where $\xi_{\rm c}(t)=R_0/c_{\rm s}t$ corresponds to the ions  passing by the target center at time $t$.
The azimuthal ion current, $J_{\rm i \phi}=Zen_{\rm i} v_{\rm i\phi}$,  is then given with the help of Eqs. (\ref{SSM1}) and (\ref{SSM2}) by
\begin{eqnarray}
\label{Jip=}
J_{\rm i\phi}(r,t)&=& \left(\frac{R_{\rm H}}{r}\right)^{2} J_{\rm H}(t),\quad r\ge R_{\rm H},
\\ 
\label{Jmax=} 
J_{\rm H}(t)&=& (R_0/R_{\rm H})Ze n_{\rm i0} c_{\rm s} (\xi_{\rm c}+1){\rm e}^{-\xi_{\rm c}-1}.
\end{eqnarray}
The spatial profile of the ion current has a maximum  $J_{\rm H}(t)$  at the Larmor hole rim  $r=R_{\rm H}$  to spatially decay at the rate $r^{-2}$ for $r\ge R_{\rm H}$.
 
In Eq. (\ref{Jmax=}), the factor $R_0/R_{\rm H}$  explains the cylindrical accumulation effect.
Consequently, the ion current contribution to the central magnetic field  follows  
$B_{\rm ci}(t)=(4\pi/c) \int^\infty_{R_{\rm H}} J_{\rm i\phi}(r,t) {\rm d}r = (4\pi/c) R_{\rm H} J_{\rm H}(t)$. 
The space-integrated quantity $B_{\rm ci}(t)$ is   independent of $R_{\rm H}$ itself. 
With time $t$, the numerical factor, $(\xi_{\rm c}+1){\rm e}^{-\xi_{\rm c}-1}$, in Eq. (\ref{Jmax=}) monotonically increases, and $\xi_{\rm c}(t)$ decreases from its initial value $\xi_{\rm c}(\tau=70$  fs) $=\xi_{\rm f}=5.5$.
Although the factor asymptotically approaches its maximum, ${\rm e}^{-1}=0.37$, with $t\rightarrow\infty$ or
$\xi_{\rm c}\rightarrow 0$, a cut-off value of $\xi_{\rm c}$ exists for the physical reason described below.

According to the EPOCH simulations, after the cavity collapse at $\tau=\tau_{\rm c}(\simeq 70$ fs),
the magnetic field  grows at the center for a period of $4-6$ 
ion-oscillations, i.e., $\Delta \tau\sim  (4-6) \times \tau_{\rm cyc} \sim 40 - 60$ fs with the cycle $\tau_{\rm cyc}=2\pi/\omega_{\rm pi}\approx$ 10 fs for the case in Figs. \ref{fig2} and \ref{fig3}. 
This corresponds to $\tau \sim  110-130$ fs or $\xi_{\rm c}\sim$ 3 (Fig. \ref{fig4}b).
After the duration $\Delta \tau$, the core periodically emits outgoing density waves at a frequency $\omega_{\rm pi}$. These waves carry a portion of the central plasma energy, as seen in the double-humped structure of the ion density profile for $\tau=110-115$ fs in Fig. \ref{fig3}a.
In fact, $B_{\rm c}(t)$ begins to decay coherently with
the first emission of the density wave (Fig. \ref{fig3}a,d). 
Since the model does not consider this emission process, we estimate the maximum magnetic field by limiting the growth at the peak time 
$\tau_{\rm peak}=\tau_{\rm c}+\Delta \tau$. 
This corresponds to $\xi_{\rm c}\sim 3$. 
Recalling the observed constancy, $B_{\rm c}\simeq 4B_{\rm ci}$, leads 
to its maximum value $B_{\rm c.max}$ as
\begin{equation}
\label{Bcmax=} 
\frac{B_{\rm c.max}}{1\,{\rm MT}}=\Psi\equiv  \frac{(Z/6)^{3/2}}{(A/12)^{1/2}} \left(\frac{n_{\rm i0}}{10^{23}\,{\rm cm}^{-3}}\right) \left(\frac{R_0}{3\,\mu {\rm m}}\right) \sqrt{\frac{\mathcal{E}_{\rm he.av}}{6\,{\rm MeV}}}.
\end{equation}
Thus, $B_{\rm c.max}$ is proportional to the total ion flux emitted from the inner surface of the microtube, i.e., $B_{\rm c.max}\propto Zn_{\rm i0}c_{\rm s}R_0$  
(recall $c_{\rm s}\propto \sqrt{Z T_{\rm e}/A}$).

\begin{figure}
\centering
\includegraphics[width=120mm]{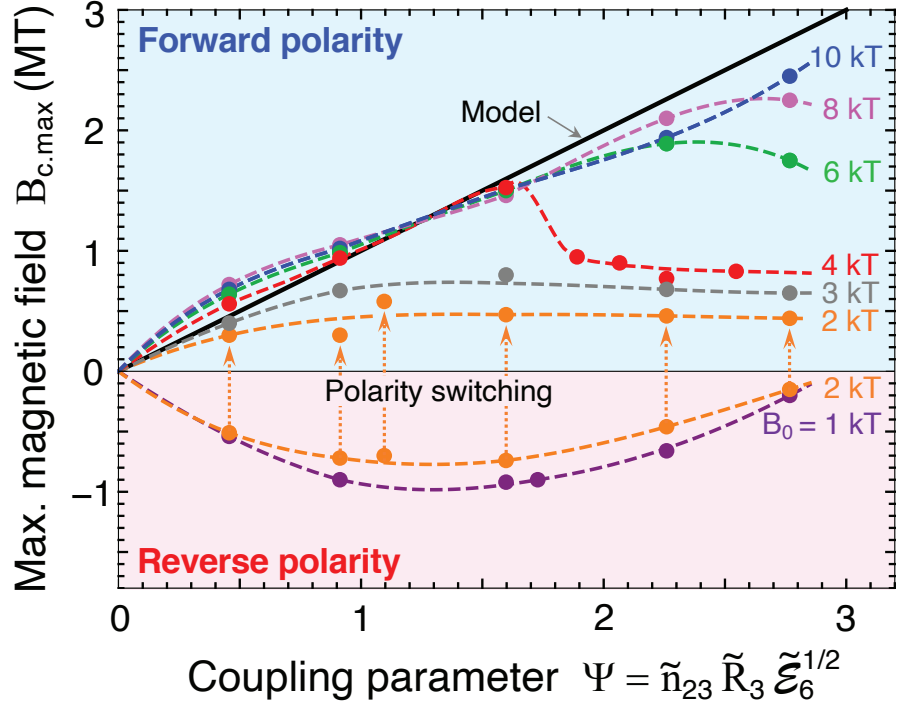}
\caption{Maximum magnetic field as a function of coupling parameter $\Psi$ [Eq. (\ref{Bcmax=})], where $\tilde{n}_{23}=n_{\rm i0}/10^{23}$cm$^{-3}$, $\tilde{R}_3=R_0/3\mu{\rm m}$, and $\tilde{\mathcal{E}}_{6}=\mathcal{E}_{\rm he.av}/6$MeV
denote normalized values for the initial ion density,  inner radius of the microtube, and average electron energy, 
respectively. 
$Z=6$ and $A=12$ are fixed assuming a fully ionized carbon plasma.
Solid circles are EPOCH results, which are linked by the dashed curves to smoothly guide the readers' eye.}
\label{fig5}
\end{figure}

Figure \ref{fig5} shows the results of about three dozen EPOCH simulations (solid circles) for $B_{\rm c.max}$ as a function of  $\Psi$ defined in Eq. (\ref{Bcmax=}).
The straight black line denotes the model prediction.
Each simulation result corresponds to a subset with the key parameters, $(B_0, n_{\rm i0}, R_0, \mathcal{E}_{\rm he.av})$. Their composition is chosen rather randomly over the ranges, $B_0=1-10$ kT, $n_{\rm i0}= 5\times 10^{22}-2\times 10^{23}$ cm$^{-3}$, $R_0=1-3\,\mu$m, and  $\mathcal{E}_{\rm he.av}=5 - 15$ MeV,  while $A=12$ and $Z=6$ are fixed. 
Despite the random choices,  the overall simulation results are smoothly linked in a systematic manner by the dashed curves parameterized by $B_0$.
This demonstrates the physical significance of the parameter $\Psi$ as an essential measure of the  magnetic-field generation  in the present scheme.
 
There is a  threshold relation between $B_0$ and $\Psi$
such that the simulation results and the model line agree well. 
For example, for $B_0\gtrsim 6$ kT and $\Psi\lesssim 2.5$, the model well reproduces the overall behavior of the simulation results. 
The physical reason why these curves overlap with the model line is as follows. 
Although the individual trajectories of charged particles are deformed to radially shift outward due to higher $B_0$ and smaller Larmor radius $R_{\rm L}$ (Fig. \ref{fig4}a), the integrated currents and consequently the magnetic field, $\int J_{\rm i\phi}{\rm d}r \propto \int J_{\rm e\phi}{\rm d}r \propto B_{\rm c}$, are  unchanged.

With $B_0=4$ kT, the difference between the simulation and the model begins to increase for $\Psi\gtrsim 1.6$.  
Moreover, for  $B_0<3$ kT, the behaviors of the simulation results are unpredictable by the model. 
In particular, with $B_0=$ 2 kT, the temporal evolution of the system becomes unstable.
Although it initially behaves in the reverse polarity regime ($B_{\rm c.max}<0$), it eventually turns into 
the forward polarity regime ($B_{\rm c.max}>0$). This phenomenon is labeled as ``polarity switching" in Fig. \ref{fig5}.
The polarity switching suddenly occurs on the timescale of several femtoseconds, when the electron current distribution surrounding the target center evolves very quickly in a complex manner.
One of the potential causes for this phenomenon seems to be the existence of an electron-rich space at the center, which is omitted in the model  assuming that $R_{\rm H}$ is so large [Eq. (\ref{RH=})] that electrons are evacuated from the Larmor hole.

\subsection*{Practical simulation with laser-plasma interaction}

{\color{black}In this section, we perform proof-of-principle EPOCH simulations to demonstrate that strong magnetic fields can still be produced, when the uniform hot electron population  previously assumed by the M-J distribution is replaced with a realistic laser-plasma interaction \cite{Arbe15}. These simulations reproduce salient features of the MTI process described in the previous sections. 
}

We consider an isolated target, which consists of an initially cold, fully ionized charge-neutral carbon$-$electron plasma with an initial
ion density $n_{\rm i0}=3\times 10^{22}$ cm$^{-3}$ 
and electron density $n_{\rm e0}=6 n_{\rm i0}$. We have reduced the target density by
a factor of $\sim 3$ relative to the case given in Figs. 2 and 3 to mitigate the computational
cost associated with PIC simulations of the laser$-$plasma interaction. The target's outer cross-section is a
square with 12-$\mu$m-long sides. The inner radius of the microtube is $R_0=3\, \mu$m.
This target is irradiated on each of the four outer sides by a large-spot (spatially plane wave) laser pulse with a wavelength of $\lambda_{\rm L}=0.8 \,\mu$m,
a total duration ($\sin^2$ temporal shape in $|E|$) of $\tau_{\rm L}=100$ fs, and a peak intensity of $I_{\rm L}=10^{21}$ W/cm$^{2}$. 
The pulses are co-timed such that the peaks of all four pulses interact with the target surface simultaneously. The plasma is modeled with a resolution of 100 cells/$\mu$m and 200 particles/cell for carbon ions and 400 particles/cell for electrons. The full size of the simulation box is 22 $\mu$m $\times$ 22 $\mu$m.

In practical laser-driven configurations, the laser$-$plasma interaction causes the hot electron population driving the implosion to depart from the conditions of spatial uniformity and temperature isotropy, which are assumed in the derivation of the maximum magnetic field given in Eq. (\ref{Bcmax=}). 
{\color{black}The energetic electron spectrum has multiple energy components and differs from a single-temperature M-J distribution (Fig. 6a). Despite this departure from the conditions assumed in the previous sections, we still observe strong magnetic field generation with similar features to the single-temperature case.}
Due to the spatial  non-uniformities, strong magnetic fields with spatially varying signs can be generated in small regions of the implosion volume even without the presence of  any seed magnetic field. However, the addition of a seed magnetic field ($B_0=6$ kT) increases both the maximum amplitude of the magnetic field produced and the spatial volume over which it maintains the same sign (Fig. 6b). 

Figure 7 shows the detailed temporal evolution of the physical quantities, which are summarized in Fig. 6b. 
The magnetic field generation in the laser-driven case occurs in two stages, where the majority of the magnetic field generation occurs during the first stage. During the implosion, substantial anisotropy in the ion current crossing through the center of the microtube generates a strong magnetic field around the center ($r<0.3\,\mu$m, Stage 1 in Fig. 7a,c). In the simulations, the magnitude of this magnetic field increases by a factor of $\gtrsim 2$ upon applying $B_0=6$ kT through the MTI process (Fig. 7a). Later, this central magnetic field is further amplified by electrons undergoing $E\times B$-directed motion as the central ion population explodes outward (Stage 2 in Fig. 7a,c). 
Unlike Stage 1, which occurs over approximately 30 fs and agrees well with the MTI process described earlier,  Stage 2 can persist for well over 100 fs. Figure 7 does not capture the end of this stage due to the computational cost of these simulations. However,  simulations performed with plastic targets suggest the magnetic field can be slowly amplified over hundreds of femtoseconds. In the presence of the 6-kT seed, the first stage, which includes the MTI amplification process, generates a $\sim 100$ kT magnetic field over $r\lesssim 0.3\,\mu$m in approximately 20 fs, while the second stage amplifies this magnetic field to $\sim 120$ kT over approximately 50 fs and increases the size of the central spot to a radius of $\sim 0.5\,\mu$m.

\section*{Discussion}
{\color{black} We here briefly discuss laser requirements for MTI.
To achieve MT-order magnetic fields experimentally, a rough estimate assuming a pulse duration of  $\sim 30$ fs suggests that a laser system with a pulse energy of $0.1-1$ kJ and a total power of $10-100$ PW  is required. 
Such high-power laser performance is accessible by today's laser technology \cite{Shen18,Dans19}.
Meanwhile, fundamental studies, including proof-of-principle experiments, should be feasible using substantially smaller laser systems.}
{\color{black}Unlike for ultrathin targets with nm-scale thicknesses, MTI targets are significantly less sensitive to laser contrast due to the micron-thick wall, while beam co-timing  should be within 10 $-$ 20 fs for implosions with a timescale of $\sim 100$ fs.}

{\color{black}Detecting MT-order magnetic fields inside a plasma presents a challenge for conventional techniques that rely on charged particle sources. In anticipation of achieving ultrahigh magnetic fields, there have been efforts to develop other techniques to infer the existence of strong B-fields inside a dense plasma in use of, for example, an XFEL photon beam with Faraday rotation effect \cite{Aref16, Weng17, Wang19} and spin-polarized neutrons \cite{Kard18}.}

{\color{black}We roughly estimate the minimum number of beams $n_{\rm B}$ from a uniformity point of view. For simplicity our discussion here is limited to the cross-sectional dimensions.
Nonuniformity of the imploding ion front is directly influenced by that of hot electron density, which comprises the local electron sheath (Fig. 1b). Hot electrons produced on the laser-irradiated surface go back and forth between the target surface and the cavity wall with an in-between distance $\Delta R=R_1-R_0$, where $R_1$ is the initial target radius. This hot electron transport is regarded as a kind of random-walk diffusion process. The nonuniformity on the outer surface should   diminish on the cavity wall via this diffusion process along the lateral direction. As is well documented, the diffusion distance is given by $L_{\rm d} \sim \sqrt{N}\Delta R$, where $N\sim c\Delta t/\Delta R$ stands for the reflection number between the two surfaces during the implosion time $\Delta t$. Consequently,  $n_{\rm B}\gtrsim 2\pi R_0/2L_{\rm d}\sim 2$ (i.e., two-sided illumination) when employing for example $R_0=3\,\mu$m,  $\Delta R=2\,\mu$m, and $\Delta t\sim 50$ fs (Fig. 2).
It should be noted that  controlling both the temporal and spacial profiles  of the incident laser pulses also plays a crucial role to improve the implosion uniformity \cite{Mura92, Mura17}.}

In summary, we propose a novel concept called MTI, which produces MT-order magnetic fields using intense  laser pulses.
Key physical elements of MTI are imploding ion fluxes with quasi-relativistic speeds  and the resultant ultrahigh spin currents running around the nanometer-scale Larmor hole at the center.
The spin currents are due to the collective motion of the imploding ions and the accompanying relativistic electrons.
The pre-seeded magnetic field $B_0$ significantly influences the magnetism of the plasma.
For example, the forward (reverse) polarity appears in the domain of higher (lower) $B_0$.
Polarity switching is an extraordinary phenomenon, which requires further investigation.
The scaling law for the maximum magnetic field $B_{\rm c.max}$ is obtained as a function of $B_0$ and the total ion flux emitted from the inner surface of microtube $\Psi$. 
With the realistic laser-plasma interaction taken into account, strong magnetic field generation has been demonstrated as a proof-of-principle of MTI. 
 
\section*{Acknowledgments} M.M. was supported by the Japan Society for the Promotion of Science (JSPS). 
J.J.H. was supported by the EUROfusion grant ENR-IFE19.CCFE-01-T001-D001 and the research grant RTI2018-098801-B-I00 of the Spanish Ministry for Research. J.J.H. thankfully acknowledges the computer resources at MareNostrum and the technical support provided by the Barcelona HPC of the Spanish Supercomputing Network and the CeSViMa HPC of the UPM.
The work by S.V.B. was supported by the project High Field Initiative 
(No. CZ.02.1.01/0.0/ 0.0/15\_003/0000449) 
from the European Regional Development Fund.
The work by K.W. and A.V.A was supported by the DOE Office of Science under Grant No. DE-SC0018312 and used HPC resources of the Texas Advanced Computing Center (TACC) at the University of Texas at Austin and the Extreme Science and Engineering Discovery Environment (XSEDE)\cite{Town14}, which is supported by National Science Foundation grant number ACI-1548562. Data collaboration was supported by the SeedMe2 project \cite{Chou17} (http://dibbs.seedme.org).

\section*{Author contributions} M.M. conceived the study. 
Designing and performing the PIC simulations are contributed by J.J.H. on Figs. 2, 3, and 5, and by K.W. and A.V.A on Figs. 6 and 7.
M.M. and S.V.B.  developed the analytical model.  
M.M. wrote the paper.  
All authors reviewed the whole work, and approved the manuscript.

\begin{figure}
\centering
\includegraphics[width=120mm]{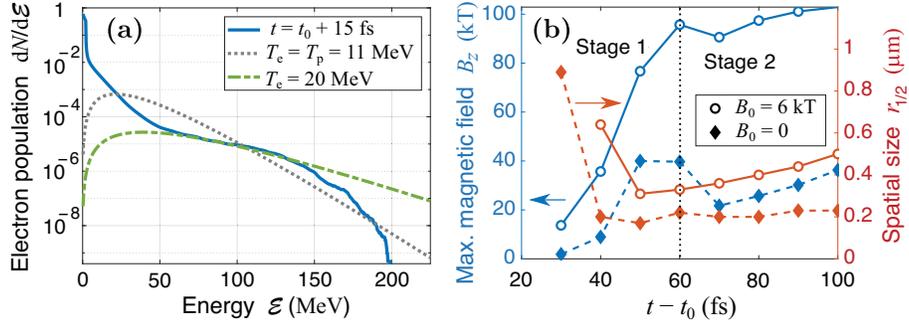}
\caption{Electron energy spectrum and magnetic field growth in laser-driven targets ($\lambda_{\rm L}= 0.8\,\mu$m, $I_{\rm L}= 10^{21}$ W/cm$^2$, $\tau_{\rm L}=100$ fs).
(a) Blue-line: Laser-produced electron energy spectrum under $B_0 = 6$ kT, normalized such that $\int ({\rm d}N/{\rm d}{\mathcal{E}}){\rm d}{\mathcal{E}} = 1$. Grey dotted line: M-J fit to the spectrum using the ponderomotive temperature $T_{\rm p} \equiv (1 + a^2_0)^{1/2}m_{\rm e}c^2 \approx 11$ MeV, where $a_0 \equiv |e|E_0/m_{\rm e}c\omega$ for a laser with maximum electric field amplitude $E_0$ and frequency $\omega$. Green dash-dotted line: M-J fit to the mid-energy part of the spectrum, provided for comparison. (b) Maximum magnetic field and radius at which the field falls to half its maximum value $(r_{1/2})$, obtained by azimuthally averaging the magnetic field;  $t_0$ is the time when the peaks of the laser pulses reach either $x = 0$ or $y = 0$.}

\label{fig6}
\end{figure}

\begin{figure}
\centering
\includegraphics[width=120mm]{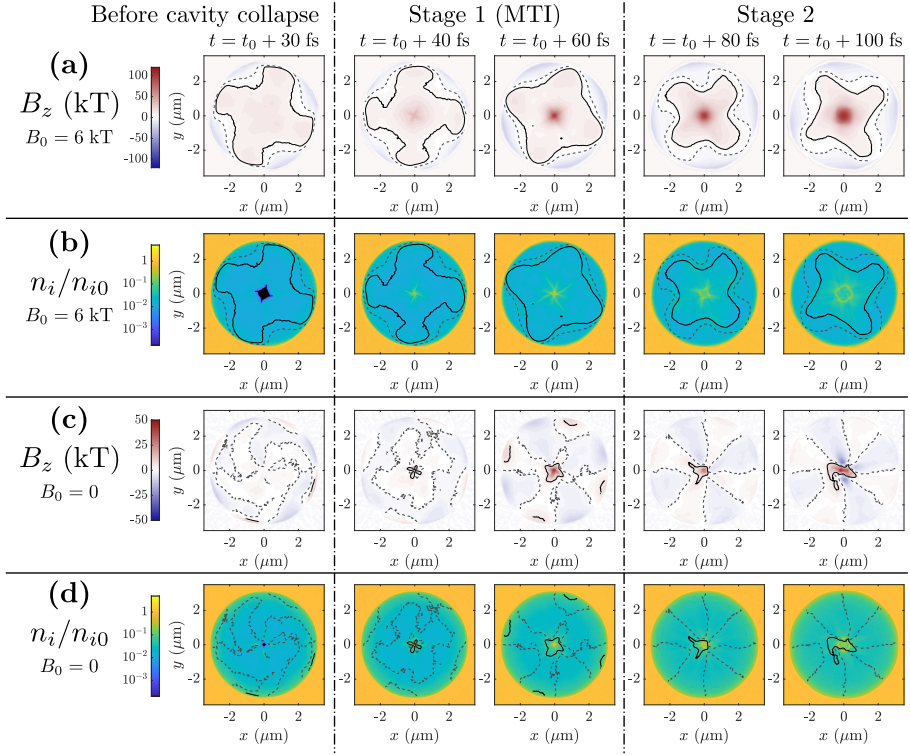}

\caption{Detailed temporal evolution of the magnetic field $B_{\rm z}$ and the normalized ion density $n_{\rm i}/n_{\rm i0}$ of laser-driven MTI. Summary plots are given in Fig. 6.
(a),(b) results with $B_0=6$ kT. (c),(d) results with $B_0=0$. (a),(c) Magnetic field in the microtube. (b),(d) Density of carbon ions during the implosion. The black solid and grey dashed contours indicate $B = 6$ kT and $B = 0$, respectively. The first time snapshot ($t=t_0+30$ fs) corresponds to a moment immediately before the cavity collapse.}

\label{fig7}
\end{figure}

\end{document}